\documentclass{llncs}

\usepackage{amsmath}
\usepackage[T1]{fontenc}
\usepackage{amsfonts}
\usepackage{amssymb}
\usepackage{float}
\usepackage{url}
\usepackage{tikz}

  \usetikzlibrary{automata}
  \usetikzlibrary{arrows}
  \usetikzlibrary{shapes}
  \usetikzlibrary{decorations.pathmorphing}
  \usetikzlibrary{fit}
  \usetikzlibrary{trees}
  \usetikzlibrary{calc}

\usepackage{algorithm}
\usepackage{algorithmic}
\usepackage{hyperref}
\usepackage[margin=1cm]{geometry}

\tikzstyle{every picture}=[
  >=stealth', shorten >=1pt, node distance=1.44cm,auto,bend angle=45,initial text=,
  every state/.style={inner sep=0.75mm, minimum size=1mm},font=\scriptsize,
  triangle/.style = { regular polygon, regular polygon sides=3, inner sep=0}
]

\newcommand{\cqfd}{\hfill{\vrule height 3pt width 5pt depth 2pt}}

\begin{document} 

  \title{Bottom Up Quotients and Residuals for Tree Languages}
  
  \author{
    Jean-Marc Champarnaud, Ludovic Mignot, Nadia Ouali-Sebti, Djelloul Ziadi
  } 

  \institute{
    LITIS, Universit\'e de Rouen, 76801 Saint-\'Etienne du Rouvray Cedex, France\\
     \email{\{Jean-Marc.Champarnaud,Ludovic.Mignot,Nadia.Ouali-Sebti,Djelloul.Ziadi\}@univ-rouen.fr}\\
  }
  
  \maketitle
  
  \begin{abstract} 
    In this paper, we extend the notion of tree language quotients to bottom-up quotients. Instead of computing the residual of a tree language from top to bottom and producing a list of tree languages, we show how to compute a set of $k$-ary trees, where $k$ is an arbitrary integer.
    We define the quotient formula for different combinations of tree languages: union, symbol products, compositions, iterated symbol products and iterated composition.    
    These computations lead to the definition of the bottom-up quotient tree automaton, that turns out to be the minimal deterministic tree automaton associated with a regular tree language in the case of the $0$-ary trees.
  \end{abstract}

\section{Introduction}\label{se:int}

Tree languages are used in numerous domains of applications in computer science,  \emph{e.g.} representation of XML documents. Regular tree languages are recognized by finite tree automata, well-studied objects leading to efficient decision problems. 
  Among them, the membership test, that is to determine whether a given word belongs to a language.
  Tree languages, that are potentially infinite, can be finitely described by regular tree expressions. Consequently it is an important subject of research to convert an expression into an equivalent automaton. 

In the case of word (that can be seen as trees with unary symbols) this is an active subject for more than fifty years: 
One of the first conversion method is the computation of the position automaton~\cite{Glu61} with linear size and a quadratic construction time w.r.t the number of occurrences of symbols in the expression.      
Three years later, Brzozowski proposed an alternative construction, the derivative automaton~\cite{Brzo64}, that is deterministic and then exponential-sized automaton. This construction is based on the operation of expression derivation, implementing the computation of the language quotient over expression. Slightly modifying this method, by replacing expression by set of expressions, Antimirov constructed the derived term automaton~\cite{Ant96} which is a linear-sized but not necessarily deterministic automaton (notice that Champarnaud and Ziadi have shown in~\cite{JMCDZ01} that this automaton is identical to Mirkin's prebase automaton~\cite{Mir66}).

Some of these methods have already been extended to tree expression: the position tree automaton was introduced in~\cite{LOZ13}, and the top-down partial derivative automaton~\cite{KM11} (see~\cite{MSZ14b,MSZ14} for an other version of the position tree automaton and its morphic links with other methods), producing non-deterministic and linear-sized tree automaton. As far as top-down deterministic tree automata are concerned, there exist regular languages that can not be recognized; Therefore, the notion of (top-down) derivative cannot be well-defined but it is not the case for bottom-up tree automata. A first step toward the computation of tree derivative as already been achieved in~\cite{CGLTT03,Lev81}, defining the bottom-up quotient of trees, that is a set of unary trees.    

In this paper, we extend the notion of bottom-up quotients to trees of any arity. Moreover, we present computation formulae for several combinations of tree languages.   
Finally, using our quotient definition,   
we present an alternative construction of the minimal bottom-up tree automaton of regular tree language \emph{via} the bottom-up quotient automaton (isomorphic to the one defined in~\cite{CGLTT03}).

\section{Preliminaries}\label{se:pre}

  See~\cite{tata} for a whole presentation about trees, tree languages and tree automata.
  
  A \emph{graded alphabet} $\Sigma=\bigcup_{k\in\mathbb{N}} \Sigma_k$ is a finite set of symbols, with $\Sigma_k$ a set of symbols of \emph{arity} $k$.
  A \emph{tree} $t$ over $\Sigma$ is inductively defined by $t=f(t_1,\ldots,t_k)$, with $k\geq 0$ any integer, $f$ any symbol in $\Sigma_k$ and $t_1$, $\ldots$, $t_k$ any $k$ trees over $\Sigma$.
  The set of the trees over $\Sigma$ is denoted by $T_\Sigma$.  
  In the following, the notion of tree is extended by considering $k$-ary trees, that are trees $k$ leaves of which are missing.
  As an example, while $f(a,g(a,b))$ is a $0$-ary tree, the tree $f(\cdot,g(a,b))$ is unary and $f(\cdot,g(\cdot,\cdot))$ is ternary.
  Given an integer $k$, $T_{\Sigma,k}$ denotes the set of the $k$-ary trees over the graded alphabet $\Sigma$.
  
  The \emph{composition} $\circ$ of trees is the operation from $T_{\Sigma,k}\times T_{\Sigma,i_1}\times\cdots\times T_{\Sigma,i_k}$ to $T_{\Sigma,i_1+ \cdots i_k}$ defined for any $k+1$ trees $t$, $t_1$, $\ldots$, $t_k$, with $t\in T_{\Sigma,k}$, denoted by $t\circ(t_1,\ldots t_k)$, as the action of grafting $t_i$ to the $i$-th missing leaf in $t$.
  Notice that $\circ$ endows $T_\Sigma$ with a structure of operad\footnote{We do not recall the different properties of operads; see~\cite{LMN13} for some examples of operads in automaton theory.}, with $()$ as an identity unary element.  
  As an example, $f(\cdot,g(\cdot,\cdot))\circ(a,b,a)=f(a,g(b,a))$.  
  To improve readability, the identity unary tree can be denoted by $\varepsilon$ (\emph{e.g.} $f(\cdot,g(\cdot,b))=f(\varepsilon,g(\varepsilon,b))$); Thus, for any $k$-ary tree $t$, $\varepsilon\circ(t)=t=t\circ(\varepsilon,\ldots,\varepsilon)$.
  
  The symbol $\varepsilon$ in a $k$-ary tree can be replaced by occurrences of distinct symbols $\varepsilon_{x_1}$, $\ldots$, $\varepsilon_{x_k}$, where $x_1$, $\ldots$, $x_k$ are any $k$ integers in $\mathbb{N}\setminus\{0\}$.
  For a $k$-ary tree $t$, we denote by $\mathrm{Ind}_\varepsilon(t)$ the set $\{x_1,\ldots,x_k\}$ of $\varepsilon$-indices. This finite and naturally ordered subsets of $\mathbb{N}$ contains the indices $x_l$ of tree symbols $\varepsilon_{x_l}$ appearing in $t$.
  Thus, a $k$-ary tree $t$ with $R$ as $\varepsilon$-indices set is inductively defined by   
  $t=\varepsilon_j$ with $j$ an integer and $k=1$, or $t=f(t_1,\ldots,t_l)$ with $f$ a symbol in $\Sigma_l$ and for $1\leq j\leq l$, $t_j$ a $n_j$-ary tree of $\varepsilon$-indices $R_j$, such that for $1\leq j,j'\leq l$, $R_j\cap R_{j'}=\emptyset$ and $\sum_{1\leq j\leq l} n_j=k$.
  In this case, the composition $t\circ(t'_1,\ldots,t'_k)$ substitutes $t'_l$ to $\varepsilon_{x_l}$ in $t$, where $\mathrm{Ind}_\varepsilon(t)=\{x_1,\ldots,x_k\}$.  
  Notice that any $k$-ary tree satisfies $\mathrm{Card}(\mathrm{Ind}_\varepsilon(t))=k$, that the occurrences are not necessarily indexed w.r.t. their apparition order in $t$, \emph{i.e.} $f(\varepsilon_2,g(\varepsilon_1,\varepsilon_3))\circ(a,b,a)=f(b,g(a,a))$ and that the empty trees are not identity elements anymore, since they may change the index of an empty tree.
  
  The composition $\circ$ is inductively defined as follows: For any $m$-ary tree $t=f(t_1,\ldots,t_n)$ with $\mathrm{Ind}_\varepsilon(t)=\{x_1,\ldots,x_m\}$, for any $m$ trees $t'_1$, $\ldots$, $t'_m$, it holds:
  \begin{align}
    \varepsilon_1\circ(t'_1) &= t'_1, &
    f(t_1,\ldots,t_n)\circ(t'_1,\ldots,t'_m)&=f((t_j\circ(t'_k)_{x_k\in\mathrm{Ind}_\varepsilon(t_j)})_{1\leq j\leq n})
   \label{eq calc ind compo} 
  \end{align}  
  
  A \emph{tree language over} $\Sigma$ is a subset of $T_\Sigma$.
  A tree language is \emph{homogeneous} if all the trees it contains admit the same arity with the same set of $\varepsilon$-indices, and $k$-\emph{homogeneous} if it only contains $k$-ary trees with the same set of $\varepsilon$-indices.
  In this case, we denote by $\mathrm{Ind}_\varepsilon(L)$ this set.
  The set of the tree languages over $\Sigma$ is denoted by $\mathcal{L}(\Sigma)$, and the set of the $k$-homogeneous tree languages by $\mathcal{L}(\Sigma)_k$ for some integer $k\geq 0$.
  Notice that the union of two $k$-homogeneous tree languages of $\varepsilon$-indices $R$ is a $k$-homogeneous tree language of $\varepsilon$-indices $R$.
  
  The composition $\circ$ is extended to an operation from $\mathcal{L}(\Sigma)_k\times (\mathcal{L}(\Sigma))^k$ to $\mathcal{L}(\Sigma)$: for any language $L$ in $\mathcal{L}_k$, for any $k$ languages $L_1$, $\ldots$, $L_k$ in $\mathcal{L}(\Sigma)$ such that $\mathrm{Ind}_\varepsilon(L_i)\cap\mathrm{Ind}_\varepsilon(L_j)=\emptyset$ for any $1\leq i<j\leq k$, $L\circ (L_1,\ldots,L_k)=\{t\circ(t_1,\ldots,t_k)\mid t\in L, t_i\in L_i, i\leq k\}$.  
  Notice that if $L_j$ is $l_j$-homogeneous for any integer $1\leq j\leq k$, then $L\circ (L_1,\ldots,L_k)$ is $\sum_{1\leq j\leq k} l_j$-homogeneous.   
  Given a $1$-homogeneous tree language $L$ of $\varepsilon$-index $\{x\}$ and an integer $n$, the \emph{iterated composition} $n_\circ$ is recursively defined by $L^{0_\circ}=\{\varepsilon_x\}$, $L^{n+1_\circ}=L^{n_\circ}\cup L^{n_\circ}\circ L$. 
  The  \emph{composition closure} of $L$ is the language $L^\circledast=\bigcup_{k\geq 0} L^{k_\circ}$. 
  Notice that $L^\circledast$ is $1$-homogeneous of $\varepsilon$-index $\{x\}$.
  
  Let $a$ be a symbol in $\Sigma_0$, $L$ be a tree language in $\mathcal{L}(\Sigma)_0$ and $t$ be a tree in $T_\Sigma$.
  The \emph{tree substitution} of $a$ by $L$ in $t$, denoted by $t_{\{a \leftarrow L\}}$, is the tree language inductively defined by
$L$ if $t=a$; 
$\{d\}$ if $t=d \in \Sigma_0\setminus\{a\}$; 
$f({t_1}_{\{a \leftarrow L\}}, \ldots, {t_k}_{\{a \leftarrow L\}})$ if $t=f(t_1, \ldots, t_k)$ with $f\in\Sigma_k$ and $t_1, \dots, t_k$ any $k$ trees over $\Sigma$.
  The $a$-\emph{product} $L_1\cdot_{a} L_2$ of two tree languages $L_1$ and $L_2$ over $\Sigma$, with $L_2$ in $\mathcal{L}(\Sigma)_0$, is the tree language $L_1\cdot_{a} L_2$ defined by $\bigcup_{t\in L_1}t_{\{a \leftarrow L_2\}}$.
  Notice that if $L_1$ is $k$-homogeneous of $\varepsilon$-index $R$, since $L_2$ is $0$-homogeneous, then $L_1\cdot_a L_2$ is $k$-homogeneous of $\varepsilon$-index $R$.  
  The \emph{iterated $a$-product} of a $0$-homogeneous tree language $L$ over $\Sigma$ is the tree language $L^{n_a}$ recursively defined by:
$L^{0_a}=\{a\}$,
$L^{{(n+1)}_a}=L^{n_a}\cup L\cdot_{a} L^{n_a}$.
  The $a$-\emph{closure} of the tree language $L$ is the language $L^{*_a}$ defined by $\bigcup_{n\geq 0} L^{n_a}$. 
  Notice that since $L$ is $0$-homogeneous, $L^{*_a}$ is $0$-homogeneous.
  
  The set of \emph{regular languages} $\mathrm{Reg}(\Sigma)$ over $\Sigma$ is the smallest set containing any subset of $\Sigma_k\circ (\Sigma_0)^k$ that is closed under union, symbol products and closures. 
  A $0$-homogeneous regular language is said to be \emph{regular} if it belongs to $\mathrm{Reg}(\Sigma)$. 
  
  Notice that the composition and the composition closure can be reinterpreted in terms of symbol product and closures.
  However, this equivalence leads to enlarge the cardinal of the alphabets and the number of operations.
  Consequently we use these operators as syntactic operations.  
  
  A \emph{tree automaton} is a $4$-tuple $A=(\Sigma,Q,F,\Delta)$ with $\Sigma$ a graded alphabet, $Q$ a set of states, $F\subset Q$ the set of final states, and $\delta$ the transition function from $\Sigma_k\times Q^k$ to $2^Q$.  
  The domain of this function can be extended to  $\Sigma_k\times (2^Q)^k$ as follows: for any symbol $f$ in $\Sigma_k$, for any $k$ subsets $Q_1$, $\ldots$, $Q_k$ of $Q$, $\delta(f,Q_1,\dots,Q_k)=\bigcup_{(q_1,\dots,q_k)\in Q_1\times\dots\times Q_k} \delta(f,q_1,\dots,q_k)$. 
  Finally, we denote by $\Delta$ the function from  $T_{\Sigma,0}$ to $2^Q$ defined for any tree in $T_{\Sigma,0}$ by 
  \begin{align*}
	  \Delta(t)&=
	    \begin{cases}
	      \delta(a) & \text{ if } t=a\in\Sigma_0,\\
	      \delta(f,\Delta(t_1),\dots,\Delta(t_k)) & \text{ if } t=f(t_1,\dots,t_k)\land f\in {\Sigma}_k \land t_1,\ldots,t_k\in T_{\Sigma,0}.
	    \end{cases}
  \end{align*}    
  A tree is \emph{accepted} by $A$ if and only if $\Delta(t)\cap F\neq \emptyset$.
  The \emph{language recognized} by $A$ is the set $L(A)$ of trees accepted by $A$, \emph{i.e.} $L(A)=\{t\in T_{\Sigma,0}\mid \Delta(t)\cap F\neq \emptyset\}$.  
  It can be shown~\cite{tata} that a tree language is recognized by some automaton if and only if it is regular.    
  A state $q$ in $Q$ is \emph{accessible} if there exists a tree $t$ in $T_\Sigma$ such that $q\in \Delta(t)$.
  Consequently, if a state is not accessible, it can be removed, and the transitions this state is a destination or a source of too, without modifying the recognized language
  A tree automaton is \emph{accessible} if all of its states are accessible.  
  A tree automaton is \emph{deterministic} if for any symbol $f$ in $\Sigma_k$, for any $k$-tuple $(q_1,\ldots,q_k)$ of states, $\mathrm{Card}(\delta(f,a_1,\ldots,q_k))\leq 1$.
  Hence, an accessible tree automaton is deterministic if and only if for any tree $t$ in $T_{\Sigma,0}$, $\mathrm{Card}(\Delta(t))\leq 1$.
  For any tree automaton $A$, there exists a deterministic tree automaton $A'$ such that $L(A)=L(A')$. 
  The automaton $A'$ can be computed from $A'$ using a subset construction~\cite{tata,RS59}.  
  The domain of the function $\Delta$ is extended to $T_{\Sigma,1}\times Q$ as follows: for any tree $t$ in $T_{\Sigma,1}$, for any state $q$ in $Q$,
  \begin{align*}
	  \Delta(t,q)&=
	    \begin{cases}
	      \{q\} & \text{ if } t=\varepsilon,\\
	      \Delta(f,Q_1,\ldots,Q_k,) & \text{ if }t=f(t_1,\ldots,t_k),
	    \end{cases}\\
	    \text{ where }\forall j\leq k, 
        Q_j&=
          \begin{cases}
            \Delta(t_j) & \text{ if }t_j\in T_{\Sigma,0},\\
            \Delta(t_j,q) & \text{ if } t_j\in T_{\Sigma,1}.
          \end{cases}
  \end{align*}  

\section{Bottom-Up Quotients} 
  In this section, we define the bottom-up quotient of a tree language w.r.t. a tree, that is an operation that delete some internal nodes in trees.
  The remaining part of the tree is usually called a context in the literature~\cite{tata}; here, we call these objects $k$-ary trees, since we need to consider the parameter $k$.
  Consequently, we reinterpret classical results from the quotient point of view.  
  Basically, the quotient is the dual operation of the composition: the quotient of a tree $t$ w.r.t. a tree $t'$ is the operation producing some trees $t''$ containing an occurrence of $\varepsilon_1$ and such that substituting $\varepsilon_1$ by $t'$ in $t''$ produces $t$. As a direct consequence, since $\varepsilon_1$ may appear in $t$, the production of $t''$ needs a reindexing of the $\varepsilon$-indices to be performed. In the following, we choose to increment these indices.   
  \begin{example}
    Let $t=f(g(\varepsilon_3,b),\varepsilon_1,h(g(\varepsilon_3,b)))$ be a tree over $\Sigma$, with $b\in\Sigma_0$, $h\in\Sigma_1$, $g\in\Sigma_2$ and $f\in\Sigma_3$. Let $t'=g(\varepsilon_3,b)$. Then
      $t'^{-1}(t)=\{f(\varepsilon_1,\varepsilon_2,h(g(\varepsilon_4,b))),$
      $f(g(\varepsilon_4,b),\varepsilon_2,h(\varepsilon_1)) \}  $.
    Indeed, it can be shown that for any $t''$ in $t'^{-1}(t)$, $t''\circ(t',\varepsilon_1,\varepsilon_3)=t$. 
  \end{example}  
  Let us formalize the notion of quotient:
  Let $t$ be a $k$-ary tree in $T_\Sigma$ and $t'$ be a $k'$-ary tree in $T_\Sigma$ such that $\mathrm{Ind}_\varepsilon(t')\subset \mathrm{Ind}_\varepsilon(t)$. 
  Let $R=\mathrm{Ind}_\varepsilon(t)$, $R'=\mathrm{Ind}_\varepsilon(t')$.
  Let $R''=\{(x_z)_{1\leq z\leq k'-k}\}=R\setminus R'$.
  The \emph{quotient of} $t$ w.r.t. $t'$ is the $k+k'-1$-homogeneous tree language $t'^{-1}(t)$ that contains all the trees $t''$ satisfying the two following conditions:
  \begin{align}
     t= t''\circ(t',(\varepsilon_{x_z})_{1\leq z\leq k-k'}),
     \quad
      \mathrm{Ind}_\varepsilon(t'')=\{1,(x_z+1)_{1\leq z\leq k'-k}\}\}\label{eq def quot tree}
  \end{align}
  As a direct consequence,
  \begin{align}
    \varepsilon_j^{-1}(\varepsilon_l)&=\{\varepsilon_1\mid j=l\} \label{eq def quot eps}\\
    t^{-1}(t')=\{\varepsilon_1\} & \Leftrightarrow t=t' \label{eq def quot t par t}
  \end{align}
  \begin{definition}\label{def quot lang}
    Let $\Sigma$ be a graded alphabet.
    Let $L$ be a tree language in $\mathcal{L}(\Sigma)$ and $t$ be a tree in $T_{\Sigma}$.
    The \emph{bottom-up quotient of} $L$ w.r.t. $t$ is the tree language $t^{-1}(L)=\bigcup_{t'\in L} t^{-1}(t')$.
  \end{definition}
  As a direct consequence of Equation~\eqref{eq def quot t par t}, the membership of a tree in a tree language can be restated in term of quotient:  
  \begin{proposition}
    Let $L$ be a tree language over a graded alphabet $\Sigma$ and $t$ be a tree in $T_{\Sigma}$.
    Then
      $t\in L \Leftrightarrow \varepsilon_1\in t^{-1}(L)$.
  \end{proposition}  
  \subsection{Bottom-Up Quotient Inductive Formulas for Trees}   
  Given a $k$-ary tree $t$ and an integer $z$, we denote by $\mathrm{Inc}_\varepsilon(z,t)$ the substitution of any symbol $\varepsilon_x$ by the symbol $\varepsilon_{x+z}$.
  Given a tree language $L$ and an integer $z$, we denote by $\mathrm{Inc}_\varepsilon(z,L)$ the tree language $\{\mathrm{Inc}_\varepsilon(z,t)\mid t\in L\}$.  
  As a direct property, it holds:
  \begin{align}
    (\mathrm{Inc}_\varepsilon(1,t))\circ((\varepsilon_{j})_{j\in\mathrm{Ind}_\varepsilon(t)}) &=t
    \label{eq inc var}
  \end{align}  
  The inductive computation of the bottom up quotient of a tree w.r.t. another tree can be performed using two basic computations:
   The bottom up quotient of a tree w.r.t. an empty tree;
   Then, the bottom up quotient of a tree w.r.t. a symbol in $\Sigma$.
  The bottom-up quotient of a tree w.r.t. a symbol of an alphabet can be inductively computed as follows: since the quotient is the inverse operation of the composition, computing the quotient  of a tree $t$ w.r.t. a tree $t'$ is in fact substituting an occurrence of $t'$ in $t$ by $\varepsilon_1$ (and increasing the $\varepsilon$-indices),
  where $t'_j=\mathrm{Inc}_\varepsilon(1,t_j)$ if $j\neq i$ and $t'_i=t'^{-1}(t_i)$.
  \begin{proposition}\label{prop calc quot symb}
    Let $\Sigma$ be a graded alphabet, 
    $k$ be an integer,
	  and $\alpha$ be a symbol in $\Sigma_k$.
	  Then:
	  \begin{align*}
	    \alpha^{-1}(\varepsilon_x)&=\emptyset, \quad
	    f^{-1}(f(\varepsilon_1,\ldots,\varepsilon_n)) = \{\varepsilon_1\},\\
	    \alpha^{-1}(f(t_1,\ldots,t_n)) &=\bigcup_{1\leq j\leq n} f(t'_1,\ldots,t'_{j-1},\alpha^{-1}(t_j),t'_{j+1},\ldots,t'_n)
	  \end{align*}
	  with $x$ an integer, $f$ a symbol in $\Sigma_n$, $t_1,\ldots,t_n$ any $n$ trees in $T_\Sigma$ and for all $1\leq z\leq n$, $t'_z=\mathrm{Inc}_\varepsilon(1,t_{z})$.
	\end{proposition}
	According to the definition of the bottom-up quotient (Equation~\eqref{eq def quot tree}) and from Definition~\ref{def quot lang}, quotienting by an indexed $\varepsilon$ is a reindexing of all the indexed $\varepsilon$.  
  \begin{proposition}\label{prop bot up quot eps}
    Let $\Sigma$ be a graded alphabet.
    Let $L$ be a $k$-homogeneous language with $\mathrm{Ind}_\varepsilon(L)=\{j_1,\ldots,j_k\}$.
    Let $j$ be an integer.
    Then:
    \begin{align*}  
      \varepsilon_j^{-1}(L)&= 
        \begin{cases}
          L\circ (
            \varepsilon_{j_1+1},\ldots,\varepsilon_{j_{z-1}+1},
            \varepsilon_1,
            \varepsilon_{j_{z+1}+1},\ldots,\varepsilon_{j_{k}+1})
          & \text{ if } j=j_z\in\mathrm{Ind}_\varepsilon(L),\\
          \emptyset & \text{otherwise.}
        \end{cases}
    \end{align*}
  \end{proposition}	
	Finally, bottom-up quotienting a tree $t$ w.r.t. to a $0$-ary tree $f(t_1,\ldots,t_k)$ can be inductively performed as follows: first, the quotient $U_k$ of $t$ w.r.t. $t_k$ is computed, producing a set of trees in which the substitution of $\varepsilon_1$ by $t_k$ produces $t$. Then the quotient $U_{k-1}$ of $U_k$ w.r.t. $t_{k-1}$ is computed, producing a set of trees in which the substitution of $\varepsilon_2$ by $t_k$ and of $\varepsilon_1$ by $t_{k-1}$ produces $t$. Eventually, the quotient $U_{1}$ of $U_2$ w.r.t. $t_{1}$ is computed, producing a set of trees in which the substitution of $\varepsilon_k$ by $t_k$, $\ldots$, and of $\varepsilon_1$ by $t_{1}$ produces $t$. Finally, the quotient $V$ of $U_1$ w.r.t. $f$ is computed, producing a set $V$ of trees in which the substitution of $\varepsilon_1$ by $f(\varepsilon_1,\ldots,\varepsilon_k)$ produces a tree in which the substitution of $\varepsilon_k$ by $t_k$, $\ldots$, and of $\varepsilon_1$ by $t_{1}$ produces $t$; therefore $V=f(t_1,\ldots,t_k)^{-1}(t)$. 
	Notice that dealing with $\varepsilon$ implies that a reindexation of the indices have to be done:
	
	If $t$ contains an occurrence of an empty tree, then its index is increased $k+1$ times by $1$, by the $k+1$ quotients; consequently, in order to quotient w.r.t. $f(t_1,\ldots,t_k)$, if an occurence of $\varepsilon_j$ appears in $t$, then the set $V$ resulting from quotienting $U_1$ by $f$ contains  some tree with an occurrence of $\varepsilon_{j+k+1}$, that has to be reindexed into $\varepsilon_{j+1}$;
	
	If $f(t_1,\ldots,t_k)$ contains an empty tree, $\varepsilon_j$ appearing in $t_l$ for example, then the set $U_{l+1}$, containing the empty trees $(\varepsilon_1,\ldots,\varepsilon_{k-l})$ (if $t$ contains some occurrences of $(t_{l+1},\ldots,t_k)$) and the empty tree $\varepsilon_{j+k-l}$, must not be quotiented w.r.t. $t_l$: if $t_l$ appears in $t$, then its $\varepsilon$ indices has been increased, and therefore $\mathrm{Inc}_\varepsilon(k-l,t_l)$ has to be considered for quotienting $U_{l+1}$.
	More formally, it can be shown that:
	\begin{proposition}\label{prop quot arbre wrt arbre}
	  Let $\Sigma$ be a graded alphabet.
	  Let $t=f(t_1,\ldots,t_k)$ be a $l$-ary tree in $T_\Sigma$ with $f$ a symbol in $\Sigma_k$ and $(t_1,\ldots,t_k)$ a $k$-tuple of trees in $T_\Sigma$ different from $(\varepsilon_1,\ldots,\varepsilon_k)$.
	  Let $u$ be a tree in $T_\Sigma$ with $\mathrm{Ind}_\varepsilon(u)=\{x_1,\ldots,x_n\}$.
	  Let $\{y_1,\ldots,y_{n-l}\}=\mathrm{Ind}_\varepsilon(u)\setminus \mathrm{Ind}_\varepsilon(t)$.
	  Then:
	  \begin{align*}
	    \begin{split}
	      t^{-1}(u)&=(f^{-1}({t'_1}^{-1}(\cdots ({t'_k}^{-1}(u))\cdots))\circ(\varepsilon_1,(\varepsilon_{y_z+1})_{1\leq z \leq n-l })\\
	      & \text{ with } \forall 1\leq j\leq  k,t'_j=\mathrm{Inc}_\varepsilon(k-j,t_j)
	    \end{split}
	  \end{align*}
	\end{proposition}
	The indexation of $\varepsilon$ plays a fundamental role in our construction: it is necessary in order to satisfy the noncommutativity of the tree operad (\emph{i.e.} $f(a,b)\neq f(b,a)$).
	\begin{example}
	  Let us consider the tree $t=g(h(a),b)$.
	  Then:
	  \begin{alignat*}{3}
	    b^{-1}(t)&= \{g(h(a),\varepsilon_1)\} 
	    &\quad & & a^{-1}(t)&=\{g(h(\varepsilon_1),b)\}\\
	    a^{-1}( b^{-1}(t))&= \{g(h(\varepsilon_1),\varepsilon_2)\} 
	    &\quad & & b^{-1}(a^{-1}(t))&= \{g(h(\varepsilon_2),\varepsilon_1)\}\\
	    h(a)^{-1}(b^{-1}(t))&= h^{-1}(a^{-1}( b^{-1}(t))) &\quad & & h(b)^{-1}(a^{-1}(t))&=h^{-1}(b^{-1}(a^{-1}(t))) \\ 
	    &= h^{-1}(g(h(\varepsilon_1),\varepsilon_2))\circ(\varepsilon_1,\varepsilon_2) 
	    &\quad & &  &= g(h^{-1}(h(\varepsilon_2)),\varepsilon_2)\\
	    &= \{g(\varepsilon_1,\varepsilon_3) \circ (\varepsilon_1,\varepsilon_2)\} &\quad & & &= \emptyset\\
	    &= \{g(\varepsilon_1,\varepsilon_2)\}\\
	    g(h(a),b)^{-1}(t)&=g^{-1}(h(a)^{-1}( b^{-1}(t))))\\
	    &= g^{-1}(g(\varepsilon_1,\varepsilon_2))\\
	    &= \{\varepsilon_1\}
	  \end{alignat*}
	  Consequently,
	  \begin{align*}
	    \{\varepsilon_1\}&=(g(h(a),b))^{-1}((g(h(a),b)))&\neq &&(g(h(b),a))^{-1}((g(h(a),b)))&=\emptyset
	  \end{align*}
	\end{example}     
  \subsection{Bottom-Up Quotient Formulas for Languages Operations}    
  Let us show now how to inductively compute the bottom-up quotient of a language w.r.t. a tree.
  As a direct consequence of Definition~\ref{def quot lang}:  
  \begin{lemma}\label{lem quot union}
    Let $\Sigma$ be a graded alphabet.
    Let $t$ be a tree in $T_\Sigma$, and $L_1$ and $L_2$ be two languages over $\Sigma$.
    Then:
      $t^{-1}(L_1\cup L_2)=t^{-1}(L_1)\cup t^{-1}(L_2)$.
  \end{lemma}  
  Then, since the sum is distributive over the composition, as a direct consequence of Lemma~\ref{lem quot union} and of Proposition~\ref{prop quot arbre wrt arbre}, it holds:  
  \begin{corollary}\label{cor deriv langage arbre}
	  Let $\Sigma$ be a graded alphabet.
	  Let $t=f(t_1,\ldots,t_k)$ be a $l$-ary tree such that $f$ is a symbol in $\Sigma_k$ and $(t_1,\ldots,t_k)$ is a $k$-tuple of trees in $T_\Sigma$ different from $(\varepsilon_1,\ldots,\varepsilon_k)$.
	  Let $L$ be a $k$-homogeneous tree language over $\Sigma$ with $\mathrm{Ind}_\varepsilon(L)=\{x_1,\ldots,x_k\}$.
	  Let $\{y_1,\ldots,y_{n-l}\}=\mathrm{Ind}_\varepsilon(L)\setminus \mathrm{Ind}_\varepsilon(t)$.
	  Then:
	  \begin{align*}
	   \begin{split}
	    t^{-1}(L)=(f^{-1}({t'_1}^{-1}(\cdots ({t'_k}^{-1}(L))\cdots))\circ(\varepsilon_1,(\varepsilon_{y_z+1})_{1\leq z\leq n-l})\\
	    \text{ with } \forall 1\leq j\leq  k,t'_j=\mathrm{Inc}_\varepsilon(k-j,t_j)
	   \end{split}
	  \end{align*}
  \end{corollary}     
  Following Corollary~\ref{cor deriv langage arbre}, it remains to show how to inductively compute the bottom-up quotient of a language w.r.t. a symbol in $\Sigma$.
  In the following, we use the \emph{partial composition} $\circ_1$ define for any $k$-ary tree $t$ (resp. $k$-homogeneous language $L$) of $\varepsilon$-indices $\{j_1,\ldots,j_k\}$ with $k\geq 1$, for any tree $t'$ (resp. tree language $L'$) by:
  \begin{align*}
    t\circ_1 t'&=t\circ(t',(\varepsilon_{l})_{j_2\leq l\leq j_k})
    &
    L\circ_1 L'&=L\circ(L',(\varepsilon_{l})_{j_2\leq l\leq j_k})
  \end{align*}  
  Let us first show how to quotient a language obtained \emph{via} a symbol product from a tree.     
  Computing a $b$-product is basically replacing any occurrence of the symbol $b$ in a tree $t$ by a tree language $L$.
  Hence, quotienting $t$ by a symbol $ \alpha$ is performed following these two conditions:
  \begin{itemize}
    \item the occurrences of $\alpha$ that have to be removed by quotienting $t\cdot_b L$ may appear in $L$. 
    However, directly computing $t\cdot_b \alpha^{-1}(L)$ may produce a tree language containing trees with several occurrences of $\varepsilon_1$.
    Therefore, we have to remove first an occurrence of $b$ in $t$, by computing $b^{-1}(t)$, then considering the substitution of the other occurrences of $b$ by $L$ in $t$, and composing the newly created $\varepsilon_1$ in $c^{-1}(t)$ with the quotient of $L$: $(b^{-1}(t)\cdot_b L)\circ_1 \alpha^{-1}(L)$; 
    \item When $\alpha\neq b$, the occurrences of $\alpha$ that have to be removed by quotienting $t\cdot_b L$ may also appear in $t$.
    In this case, an occurrence of $\alpha$ has to be substituted by $\varepsilon_1$, and the occurrences of $b$ in $t$ are still replaced by $L$: $\alpha^{-1}(t)\cdot_b L$.
  \end{itemize}
  This is illustrated in the next lemma.
  \begin{lemma}\label{lem quot tree cdotb}
    Let $\Sigma$ be a graded alphabet.
    Let $t$ be a $k$-ary tree in $T_\Sigma$ and $L$ be a $0$-homogeneous language.
    Let $\alpha$ be a symbol in $\Sigma$ and $b$ be a symbol in $\Sigma_0$.
    Then:
    \begin{align*}  
      \alpha^{-1}(t\cdot_b L)&= 
        \begin{cases}
          (b^{-1}(t)\cdot_b L) \circ_1 b^{-1}(L) & \text{ if }\alpha=b,\\
          \alpha^{-1}(t)\cdot_b L \cup (b^{-1}(t)\cdot_b L) \circ_1 \alpha^{-1}(L) & \text{ if }\alpha\in\Sigma_0\setminus\{b\},\\
          \alpha^{-1}(t)\cdot_b L & \text{otherwise.}
        \end{cases}
    \end{align*}
  \end{lemma}
  Hence, as a direct consequence of Lemma~\ref{lem quot tree cdotb}, since $L\cdot_b L'=\bigcup_{t\in L} t\cdot_b L'$:
  \begin{proposition}\label{prop quot cdotb lang}
    Let $\Sigma$ be a graded alphabet.
    Let $L_1$ be a $k$-homogeneous language and $L_2$ be a $0$-homogeneous language.
    Let $\alpha$ be a symbol in $\Sigma$ and $b$ be a symbol in $\Sigma_0$.
    Then:
    \begin{align*}  
      \alpha^{-1}(L_1\cdot_b L_2)&= 
        \begin{cases}
          (b^{-1}(L_1)\cdot_b L_2) \circ_1 b^{-1}(L_2) & \text{ if }\alpha=b,\\
          \alpha^{-1}(L_1)\cdot_b L_2 \cup (b^{-1}(L_1)\cdot_b L_2) \circ_1 \alpha^{-1}(L_2) & \text{ if }\alpha\in\Sigma_0\setminus\{b\},\\
          \alpha^{-1}(L_1)\cdot_b L_2 & \text{otherwise,}\\
        \end{cases}
    \end{align*}
  \end{proposition}
  Let us now explain how to quotient a tree obtained \emph{via} the composition w.r.t. a $n$ ary symbol $\alpha$. 
  Composing a $k$-ary tree $t$, satisfying $\mathrm{Ind}_\varepsilon(t)=\{x_1,\ldots,x_k\}$, with $k$ trees $t_1,\ldots,t_k$ is the action of grasping these trees to $t$ at the positions where $\varepsilon_{x_1},\ldots,\varepsilon_{x_k}$ appear. Hence, the resulting tree $t'$ can be viewed as a tree with an upper part containing $t$ and the lower parts containing $t_1,\ldots,t_k$ exactly. 
  Then, if $\alpha$ appears in a lower tree $t_j$, this tree has to be quotiented w.r.t. $\alpha$ and the other trees are $\varepsilon$-incremented.
  Moreover, if some $n$ trees in $t_1,\ldots,t_k$ are equal to $\varepsilon_1,\ldots,\varepsilon_n$, let us say $t_{p_1},\ldots,t_{p_n}$, and if $t'=\alpha(\varepsilon_{x_{p_1}},\ldots,\varepsilon_{x_{p_n}})$ appears in $t$, then $t'$ has to be substituted by $\varepsilon_1$ and the other lower trees $t_j$ such that $j\neq p_m$, $m\in\{1,\ldots,n\}$ $\varepsilon$-incremented, since the inverse operations produce $t$.
  More formally,
  \begin{lemma}\label{lem quot tree compos}
    Let $\Sigma$ be a graded alphabet.
    Let $t$ be a $k$-ary tree with $\mathrm{Ind}_\varepsilon(t)=\{j_1,\ldots,j_k\}$  and $t_1,\ldots,t_k$ be $k$ trees.
    Let $\alpha$ be a symbol in $\Sigma_n$.
    Then:
    \begin{align*} 
      \alpha^{-1}(t\circ(t_1,\ldots,t_k))&=
          \bigcup_{1\leq j\leq k} t\circ((\mathrm{Inc}_\varepsilon(1,t_l))_{1\leq l\leq j-1},\alpha^{-1}(t_j),(\mathrm{Inc}_\varepsilon(1,t_l))_{j+1\leq l\leq k})\\
          &\quad \cup  
            \begin{cases}
               \alpha((\varepsilon_{j_{p_l}})_{1\leq l\leq n})^{-1}(t) \circ(\varepsilon_1,(\mathrm{Inc}_\varepsilon(1,t_l))_{1\leq l\leq k\mid\forall j, l\neq p_j}) \\
               \quad \text{ if }\forall 1\leq l\leq n,  \exists 1\leq p_l\leq k, t_{p_l}= \varepsilon_l, \\
               \emptyset  \quad \text{ otherwise.}
            \end{cases} 
    \end{align*}
  \end{lemma}
  Therefore, since $L\circ(L_1,\ldots,L_k)=\bigcup_{t\in L,(t_1,\ldots,t_k)\in L_1\times\cdots\times L_k} t\circ (t_1,\ldots,t_k)$:
  \begin{proposition}\label{prop quot lang circ}
    Let $\Sigma$ be a graded alphabet.
    Let $L$ be a $k$-homogeneous language with $\mathrm{Ind}_\varepsilon(L)=\{j_1,\ldots,j_k\}$ and $L_1,\ldots,L_k$ be $k$ tree languages.
    Let $\alpha$ be a symbol in $\Sigma_n$.
    Then:
    \begin{align*} 
      \alpha^{-1}(L\circ(L_1,\ldots,L_k))&=  \bigcup_{1\leq j\leq k} L\circ((\mathrm{Inc}_\varepsilon(1,L_l))_{1\leq l\leq j},\alpha^{-1}(L_j),(\mathrm{Inc}_\varepsilon(1,L_l))_{j+1\leq l\leq k}) \\
      &\quad \cup
        \begin{cases}
          \alpha((\varepsilon_{j_{p_l}})_{1\leq l\leq n})^{-1}(L) \circ(\varepsilon_1,(\mathrm{Inc}_\varepsilon(1,L_l))_{1\leq l\leq k\mid \forall z, l\neq p_z}))\\
          \quad \text{ if } \forall 1\leq l\leq n, \exists 1\leq p_l\leq k, \varepsilon_l \in L_{p_l}\\
          \emptyset \quad \text{ otherwise.}
        \end{cases}
    \end{align*}
  \end{proposition}
  Finally, the two iterated operations are quotiented as follows.
  
  The iterated composition can be quotiented by a $n$-ary symbol $\alpha$ with $n\leq 1$.
  If $n=1$, since $L^{\circledast}$ is obtained by applying an arbitrary number of times the composition, then quotienting w.r.t. $\alpha$ is quotienting tree $t$ in $L$ w.r.T. $\alpha$ and grasping it to the language obtained by an arbitrary number of application of the composition, that is $L^{\circledast}$. Equivalently, the occurrence of $\alpha$ to remove appears in a lower part of a tree.
  However, when $n=0$, the occurrence of $\alpha$ to remove can appear everywhere: it can be localized under an upper tree in $L^{\circledast}$ but above a lower tree in $L^{\circledast}$ too, that is when the tree $t$ to quotient belongs to $L^{\circledast}\circ \{t'\}\circ L^{\circledast}$ and when the occurrence of $\alpha$ to remove appears in $t'\in L$. In this case, $t'$ has to be quotiented w.r.t. $\alpha$, creating an occurrence of $\varepsilon_1$, and then the former unique $\varepsilon$-index of $L^{\circledast}$ has to be incremented, in line with the definition of the bottom up quotient. Hence:
  \begin{proposition}\label{prop bot up quot star rond}
    Let $\Sigma$ be a graded alphabet.
    Let $L$ be a $1$-homogeneous language.
    Let $\alpha$ be a symbol in $\Sigma_0$.
    Then:
    \begin{align*} 
      \alpha^{-1}(L^{\circledast})&=
        \begin{cases}
          (L^\circledast\circ (\alpha^{-1}(L)))\circ(\varepsilon_1, \mathrm{Inc}_{\varepsilon}(1,L^\circledast)) & \text{ if }\alpha\in\Sigma_0,\\
          (L^\circledast\circ (\alpha^{-1}(L))) & \text{ otherwise.}
        \end{cases}
    \end{align*}
  \end{proposition}
  In the case of the iterated $b$-product $L^{*_b}$, two cases are considered when quotienting w.r.t. $\alpha$:
  when $b=\alpha$, then one occurrence of $b$ in a tree in $L$ has to be transformed into $\varepsilon_1$, whereas the other may still be substituted by $L$.
  But when $\alpha\neq b$, then the situation is more complex: likely to the second case of the iterated composition, the occurrence $\alpha$ to be removed may appear everywhere: it can be localized under an upper tree in $L^{*_b}$ when it was substituted from an occurrence of $b$, but above a lower tree in $L^{*_b}$ too, if it also contains an occurrence of $b$. This may occurs when the tree $t$ to quotient belongs to $L^{*_b}\cdot_b \{t'\}\cdot_b L^{*_b}$ and when the occurrence of $\alpha$ to remove appears in $t'\in L$. In this case, $L$ has to be quotiented first w.r.t. $b$ in order to create a new occurrence of $\varepsilon_1$, where the quotient $\alpha^{-1}(L)$ is grasped. Then a $b$-product is added, since any occurrence of $b$ still may be substituted by $L^{*_b}$. Consequently:
  \begin{proposition}\label{prop bot up quot star symb}
    Let $\Sigma$ be a graded alphabet.
    Let $L$ be a $0$-homogeneous language.
    Let $\alpha$ be a symbol in $\Sigma$ and $b$ be a symbol in $\Sigma_0$.
    Then:
    \begin{align*} 
      \alpha^{-1}(L^{*_b})&=
        \begin{cases}
          (b^{-1}(L))^\circledast\cdot_b L^{*_b} & \text{ if }\alpha=b,\\
          ((b^{-1}(L))^\circledast \circ (\alpha^{-1}(L))) \cdot_b L^{*_b} & \text{otherwise.}\\
        \end{cases}
    \end{align*}
  \end{proposition}
  In the following section, we show how to make use of these quotients in order to compute the minimal tree DFA associated with a $0$-homogeneous recognizable tree language.  
  \section{The Bottom-Up Quotient Automaton}   
  Let $A=(\Sigma,Q,F,\delta)$ be a (non-necessarily deterministic) tree automaton and $q$ be a state in $Q$.
  The \emph{top language of} $q$ is $L^{q}(A)=\{t' \in T_{\Sigma,1} \mid \Delta(t,q)\cap F\neq\emptyset\}$.
  The \emph{down language} of $q$ is the tree language $L_q(a)=\{t\mid t\in T_{\Sigma,k}\land q\in \Delta(t)\}$.
  Hence, a state $q$ is accessible if and only if $L_q(A)$ is not empty.
  The bottom up quotient is related to the top language of a state as follows:
%
%
  \begin{proposition}\label{prop lien quot lang et lang haut}
    Let $A=(\Sigma,Q,F,\delta)$ be a automaton.
    Then, for any tree $t$ in $T_{\Sigma}$, it holds:
      $t^{-1}(L(A))=
        \bigcup_{q\in \Delta(t)} L^{q}(A)$.
  \end{proposition}
  Consequently, since there exists only a finite set of combination of states:
  \begin{theorem}\label{thm lien card aut card quot}
    Let $A=(\Sigma,Q,F,\delta)$ be a deterministic tree automaton.
    Then:
    \begin{align*}
      \mathrm{Card}(\{t^{-1}(L(A))\mid t\in T_{\Sigma}\}) \leq \mathrm{Card}(Q)
    \end{align*}
  \end{theorem}
  Let $L$ be a tree language in $\mathcal{L}(\Sigma)_0$.
    The \emph{bottom-up quotient automaton of} $L$ is the automaton $A_L=(\Sigma,Q,F,\delta)$ defined by:
       $Q=\{t^{-1}(L)\mid t\in T_{\Sigma}\}$,
       $F=\{L'\in Q \mid\varepsilon_1\in L'\}$,
       $\delta(f,t_1^{-1}(L),\ldots,t_k^{-1}(L))=\{f(t_1,\ldots,t_k)^{-1}(L)\}$.
    
  \begin{proposition}\label{prop aut min reco l}  
    Let $L$ be a tree language in $\mathcal{L}(\Sigma)_0$.
    Then $L(A_L)=L$.
  \end{proposition}
  Let us now show that any deterministic tree automaton recognizing a tree language $L$ can be send onto the bottom-up quotient automaton of $L$ \emph{via} a particular morphism associating any state with its top language, defined as follows:
  \begin{definition}
    Let $A_1=(\Sigma,Q_1,F_1,\delta_1)$ and $A_2=(\Sigma,Q_2,F_2,\delta_2)$ be two tree automata.
    A \emph{morphism} $\phi$ from $A_1$ to $A_2$ is a function from $Q_1$ to $Q_2$ that satisfies the two following conditions: \textbf{(1)} $\phi(F_1)\subset F_2$, \textbf{(2)} for any transition $(q,f,q_1,\ldots,q_k)$ in $\delta_1$, $(\phi(q),f,\phi(q_1),\ldots,\phi(q_k))$ is in $\delta_2$.
  \end{definition}
  
  \begin{proposition}\label{prop morphism from dfa to min}
    Let $A$ be an accessible deterministic tree automaton.
    Let $\phi$ be the function that associates to any state $q$ in $Q$ the language $L^q(A)$.
    Then $\phi$ a morphism from $A$ to $A_{L(A)}$.
  \end{proposition}
%
%
  As a direct consequence of Proposition~\ref{prop morphism from dfa to min},
  \begin{theorem}
    Let $L$ be a recognizable tree language.
    Then:
    \begin{align*}
      \text{The minimal DFA associated with $L$ is unique (up to an isomorphism) and is $A_L$.}
    \end{align*}
  \end{theorem}

\section{Example}

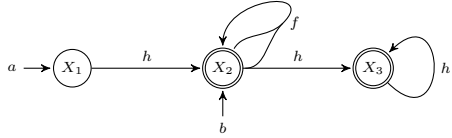
\begin{figure}[H]
  \centerline{
	\begin{tikzpicture}[node distance=2.5cm,bend angle=30,transform shape,scale=0.8]
	  \node[state] (X1) {$X_1$};
	  \node[state,accepting, right of=X1] (X2) {$X_2$};	
	  \node[state,accepting, right of=X2] (X3) {$X_3$}; 
	  \draw (X1) ++(-1cm,0cm) node {$a$}  edge[->] (X1); 
	  \draw (X2) ++(-0cm,-1cm) node {$b$}  edge[->] (X2); 
      \path[->]
        (X1) edge[->,above] node {$h$} (X2)
        (X2) edge[->,above] node {$h$} (X3)
        (X3) edge[->,in=45,out=-45,loop,right] node {$h$} ()
	  ;
      \draw (X2) ++(1cm,0.75cm)  edge[->,in=90,out=45,looseness=2] node[right,pos=0] {$f$} (X2) edge[out=225,in=60] (X2) edge[out=225,in=0] (X2);
    \end{tikzpicture}
  }
  \caption{The Minimal Tree Automaton of $L_1$.}
  \label{fig buquotDFA ex}
\end{figure}

  Let us consider an alphabet $\Sigma$ satisfying $\Sigma_0=\{a,b\}$, $\Sigma_1=\{h\}$ and $\Sigma_2=\{f\}$.
  In this section, we show how to compute the bottom-up quotient automaton of the tree language $L_1=h^{\circledast}\circ L_2$ with $L_2=(h(a)+f(b,b))^{*_b}$.
  First, let us consider $L_3=(f(\varepsilon_1,b) + f(b,\varepsilon_1))^{\circledast}$ and $L'_3=(f(\varepsilon_2,b) + f(b,\varepsilon_2))^{\circledast}$.
  \begin{align*}
    b^{-1}(L_3) &= (L_3 \circ (f(\varepsilon_1,\varepsilon_2)+f(\varepsilon_2,\varepsilon_1)))\circ(\varepsilon_1,L'_3)\\
      &= L_3 \circ (f(\varepsilon_1,L'_3)+f(L'_3,\varepsilon_1)) = L_4\\
    f^{-1}(L_4)&=L_3 \circ (f^{-1}f(\varepsilon_1,L'_3)+f^{-1}f(L'_3,\varepsilon_1)) =L_3
  \end{align*}  
  Let us now consider $L_2=(h(a)+f(b,b))^{*_b}$.
  \begin{align*}
    a^{-1}(L_2) &=(L_3\circ h(\varepsilon_1)) \cdot_b  L_2 \  & b^{-1}(b^{-1}(L_2))&=(b^{-1}(L_3)  \cdot_b  L_2)\circ_1 (b^{-1}(L_2))\\
    h(a)^{-1}(L_2)& =L_3 \cdot_b  L_2                            & &=(L_4  \cdot_b  L_2)\circ_1 (L_3\cdot_b L_2)\\
    b^{-1}(L_2)&=L_3  \cdot_b  L_2                               & &=(L_4 \circ_1 L_3) \cdot_b L_2\\
    f(b,b)^{-1}(L_2)&=L_3\cdot_b L_2                             & &=(L_3 \circ (f(L_3,L'_3)+f(L'_3,L_3)) ) \cdot_b L_2\\
  \end{align*}
  Let us consider the language $L_1=h^{\circledast}\circ L_2$.
  \begin{align*}
    a^{-1}(L)&=h^{\circledast}\circ a^{-1}(L_2) &
    h(a)^{-1}(L)&=h^{\circledast}\circ h(a)^{-1}(L_2)\\
    &=h^{\circledast}\circ ((L_3\circ h(\varepsilon_1)) \cdot_b  L_2) =X_1&
    &=h^{\circledast}\circ(L_3 \cdot_b  L_2)=X_2\\
    b^{-1}(L)&= h^{\circledast}\circ b^{-1}(L_2) &
    h(b)^{-1}(L)&=h^{\circledast}\circ h(b)^{-1}(L_2)\\
    &=   h^{\circledast}\circ(L_3  \cdot_b  L_2) =X_2&
    &=h^{\circledast} \circ(L_3 \cdot_b  L_2)=X_2\\
    & & &=h^{-1}h^{\circledast}\\
    & & &=h^{\circledast}=X_3\\
    h(h(b))^{-1}(L)&=h^{-1}h^{\circledast}&
    f(b,b)^{-1}(L)&=h^{\circledast}\circ f(b,b)^{-1}(L_2)\\
    &=h^{\circledast}=X_3&
    &=h^{\circledast}\circ( L_3\cdot_b L_2)=X_2
  \end{align*}
  The bottom up quotient of $L_1$ is given in Figure~\ref{fig buquotDFA ex}.
  
\section{Conclusion and Perspectives}
  In this paper, we have extended the notion of language quotient to tree languages bottom up quotient.
  We have also shown that using these languages, the minimal deterministic tree automaton of a recognizable tree language is obtained by merging the states that share the same top languages.  
  However, this techniques is intractable since it is defined over tree languages only.
  It could be interesting to define the notion of tree expression derivatives, due to Brzozowski~\cite{Brzo64}, that allows the computation of a deterministic automaton from a regular expression.
  This involves another theoretical interpretation of the regular tree expression, since two new operators (the composition and its iteraed version) are necessary.
  Moreover, the set of expression derivatives from an expression is not necessarily finite in the case of words, and consequently is not in the case of tree.
  Brzozowski shown that the ACI equivalence of the sum is sufficient to obtain a finite set of reduced expression: but is it the case for tree expressions ?
  
  Another step is the investigation of non-necessarily $0$-ary tree languages, since the notion of quotient is defined for any arity, as far as the homogeneity of tree languages is concerned.
  Is there exist a notion of minimal deterministic tree automaton for such languages ?

\bibliography{biblio}

\appendix
\newpage
\section{Proofs}

  \noindent\textbf{Proposition~\ref{prop calc quot symb}.}
    \emph{Let $\Sigma$ be a graded alphabet, 
    $k$ be an integer,
	  and $\alpha$ be a symbol in $\Sigma_k$.
	  Then:}
	  \begin{align*}
	    \alpha^{-1}(\varepsilon_x)&=\emptyset,\\
	    \alpha^{-1}(f(t_1,\ldots,t_n)) &=
	      \begin{cases}
	        \{\varepsilon_1\} & \text{ \emph{if} }\alpha=f\\
	        & \land \forall 1\leq z\leq n, t_z=\varepsilon_z,\\
	        \bigcup_{1\leq j\leq n} f(t'_1,\ldots,t'_{j-1},\alpha^{-1}(t_j),t'_{j+1},\ldots,t'_n) & \text{ \emph{otherwise,}}
	      \end{cases}
	  \end{align*}
	  \emph{with $x$ an integer, $f$ a symbol in $\Sigma_n$, $t_1,\ldots,t_n$ any $n$ trees in $T_\Sigma$ and for all $1\leq z\leq n$, $t'_z=\mathrm{Inc}_\varepsilon(1,t_{z})$.}
	\begin{proof}
	  Let $t$ be a $l$-ary tree $t$ in $T_\Sigma$ with $\mathrm{Ind}_\varepsilon(t)=\{x_1,\ldots,x_l\}$.
	  Obviously, if  $\{1,\ldots,k\}$ is not included into $\mathrm{Ind}_\varepsilon(t)$, then $\alpha^{-1}(t)=\emptyset$.
	  Thus suppose that $\{1,\ldots,k\}\subset\mathrm{Ind}_\varepsilon(t)$.
	  \begin{enumerate}
	    \item If $t=\varepsilon_x$, since $t\neq\varepsilon_x$, it holds from Equation~\eqref{eq def quot eps} that $\alpha^{-1}(t)=\emptyset$.
	    \item If $t=\alpha$, it holds from Equation~\eqref{eq def quot t par t} that $\alpha^{-1}(t)=\{\varepsilon_1\}$.
	    \item Consider that $t=f(t_1,\ldots,t_n)\neq\alpha$. 
	    Let us set $S=\bigcup_{1\leq j\leq n} f(t'_1,\ldots,t'_{j-1},$ $\alpha^{-1}(t_j),t'_{j+1},\ldots,t'_n)$.
	    For any integer $1\leq j\leq n$, let us define the set $\Gamma_j$ by $(\varepsilon_{x_z})_{1\leq z\leq l \land k+1\leq x_z \land x_z\in \mathrm{Ind}_\varepsilon(t_j) }$ and $\Gamma=(\varepsilon_{x_z})_{1\leq z\leq l \land k+1\leq x_z}$.
	    From Equation~\eqref{eq def quot tree}, for any integer $1\leq j\leq n$, $\alpha^{-1}(t_j)$ is the only set satisfying for any tree $v$ it contains
			  \begin{align*}
			    t_j &= v\circ(\alpha,\Gamma_j),\\
			    \mathrm{Ind}_\varepsilon(v)&=\{1\}\cup\{x+1\mid \varepsilon_{x}\in\Gamma_j\}
			  \end{align*}
			Since by construction of $S$, $v\in \alpha^{-1}(t_j)$  $\Leftrightarrow$ $f(t'_1,t'_{j-1},v,t'_{j+1},\ldots,t'_{n}) \in S$,
			$S$ is the only set satisfying for any tree $v$ it contains
			  \begin{align*}
			    t &= v\circ(\alpha,\Gamma),\\
			    \mathrm{Ind}_\varepsilon(v)&=\{1\}\cup\{x+1\mid \varepsilon_{x}\in\Gamma\}
			  \end{align*}
			that is $\alpha^{-1}(t)$.
	  \end{enumerate}
	  \cqfd
	\end{proof}	
	
	\noindent\textbf{Proposition~\ref{prop bot up quot eps}.}
    \emph{Let $\Sigma$ be a graded alphabet.
    Let $L$ be a $k$-homogeneous language with $\mathrm{Ind}_\varepsilon(L)=\{j_1,\ldots,j_k\}$.
    Let $j$ be an integer.}
    Then:
    \begin{align*}  
      \varepsilon_j^{-1}(L)&= 
        \begin{cases}
          L\circ (
            \varepsilon_{j_1+1},\ldots,\varepsilon_{j_{z-1}+1},
            \varepsilon_1,
            \varepsilon_{j_{z+1}+1},\ldots,\varepsilon_{j_{k}+1})
          & \text{ \emph{if} } j=j_z\in\mathrm{Ind}_\varepsilon(L),\\
          \emptyset & \text{\emph{otherwise}.}
        \end{cases}
    \end{align*}
  \begin{proof}
    Let us show by induction over the tree structure that for any tree $t$ in $L$,
    \begin{align*}  
      \varepsilon_j^{-1}(t)&= 
        \begin{cases}
          \{t\circ (
            \varepsilon_{j_1+1},\ldots,\varepsilon_{j_{z-1}+1},
            \varepsilon_1,
            \varepsilon_{j_{z+1}+1},\ldots,\varepsilon_{j_{k}+1}) \}
          & \text{ if } j=j_z\in\mathrm{Ind}_\varepsilon(L),\\
          \emptyset & \text{otherwise.}
        \end{cases}
    \end{align*}
    \begin{enumerate}
      \item If $t=\varepsilon_{j_1}$, it holds from Equation~\eqref{eq def quot eps} that 
		    \begin{align*}  
		      \varepsilon_j^{-1}(t)&= 
		        \begin{cases}
		            \varepsilon_1=t\circ\varepsilon_1,
		          & \text{ if } j=j_1,\\
		          \emptyset & \text{otherwise.}
		        \end{cases}
		    \end{align*}
		  \item Consider that $t=f(t_1,\ldots,t_n)$ with $f$ in $\Sigma_n$ and $t_1,\ldots,t_n$ any $n$ trees. 
		  Obviously, if $j\notin \mathrm{Ind}_\varepsilon(t)$, then it holds from Equation~\eqref{eq def quot eps} that $ \varepsilon_j^{-1}(t)=\emptyset$.
		  Thus conside that $j\in \mathrm{Ind}_\varepsilon(t)$.
		  By definition, $t$ is the only tree $v$ satisfying
		  \begin{align*}
		     t&= v\circ((\varepsilon_{j_z})_{1\leq z\leq k}),\\
		     \mathrm{Ind}_\varepsilon(v)&=\mathrm{Ind}_\varepsilon(t)
		  \end{align*}
		  Therefore $t'=t\circ ( \varepsilon_{j_1+1},\ldots,\varepsilon_{j_{z-1}+1}, \varepsilon_1, \varepsilon_{j_{z+1}+1},\ldots,\varepsilon_{j_{k}+1})$ is the only tree $v$ satisfying
		  \begin{align*}
		     t&= t'\circ((\varepsilon_{j_z})_{1\leq z\leq k}),\\
		     \mathrm{Ind}_\varepsilon(t')&=\{1,(x_z+1)_{1\leq z\leq k,x_z\neq j}\}
		  \end{align*}
		  Consequently, $\varepsilon_j^{-1}(t)=\{t'\}$ and by Definition~\ref{def quot lang}, the proposition is valid.
    \end{enumerate}
    \cqfd
  \end{proof}
  
  \noindent\textbf{Proposition~\ref{prop quot arbre wrt arbre}.}
	  \emph{Let $\Sigma$ be a graded alphabet.
	  Let $t=f(t_1,\ldots,t_k)$ be a $l$-ary tree in $T_\Sigma$ with $f$ a symbol in $\Sigma_k$ and $(t_1,\ldots,t_k)$ a $k$-tuple of trees in $T_\Sigma$ different from $(\varepsilon_1,\ldots,\varepsilon_k)$.
	  Let $u$ be a tree in $T_\Sigma$ with $\mathrm{Ind}_\varepsilon(u)=\{x_1,\ldots,x_n\}$.
	  Let $\{y_1,\ldots,y_{n-l}\}=\mathrm{Ind}_\varepsilon(u)\setminus \mathrm{Ind}_\varepsilon(t)$.
	  Then:}
	  \begin{align*}
	    \begin{split}
	      t^{-1}(u)&=(f^{-1}({t'_1}^{-1}(\cdots ({t'_k}^{-1}(u))\cdots))\circ(\varepsilon_1,(\varepsilon_{y_z+1})_{1\leq z \leq n-l })\\
	      & \text{ \emph{with} } \forall 1\leq j\leq  k,t'_j=\mathrm{Inc}_\varepsilon(k-j,t_j)
	    \end{split}
	  \end{align*}
	\begin{proof}
  For any integer $1\leq j\leq k$, let us set $U'_j=({t'_j}^{-1}(\cdots ({t'_k}^{-1}(u))\cdots)$, $\Gamma_j=(\varepsilon_{x_z})_{1\leq z \leq n\land x_z\notin \bigcup_{j\leq l\leq k}\mathrm{Ind}_\varepsilon(t_l)}$, and $\Delta_j=(\varepsilon_1,\ldots,\varepsilon_{k-j+1},\mathrm{Inc}(k-j+1,\Gamma_{j-1}))$.
  Notice that by definition, $\Gamma_1=\mathrm{Ind}_\varepsilon(u)\setminus \mathrm{Ind}_\varepsilon(t)$.
	  \begin{enumerate}
		  \item By a downward recurrence over $j$ from $k$ to $1$, and by induction over $t$, let us show that $U'_j$ is the only set satisfying $\forall v\in U'_j$,
		  \begin{align*}
		    u&=v\circ(t_j,\ldots,t_k,\Gamma_j)\\
		    \mathrm{Ind}_\varepsilon(v) &=\{1,\ldots,k-j+1\}\cup \{x+k-j+1\mid \varepsilon_x\in\Gamma_j\}\\
		    \text{(\emph{i.e.} } \mathrm{Ind}_\varepsilon(v) &= \{1\}\cup\mathrm{Inc}_\varepsilon(1,\Delta_{j+1}))
		    \intertext{When $j=k$, from Equation~\eqref{eq def quot tree}, since $U'_k=t_k^{-1}u$, $U'_k$ is the only set satisfying $\forall v\in U'_k$}
		    u&=v\circ(t_k,\Gamma_k)\\
		    \mathrm{Ind}_\varepsilon(v) &=\{1\}\cup \{x+1\mid \varepsilon_x\in\Gamma_k\}
		    \intertext{By recurrence hypothesis, $U'_j$ is the only set satisfying $\forall v\in U'_j$}
		    u&=v\circ(t_j,\ldots,t_k,\Gamma_j)\\
		    \mathrm{Ind}_\varepsilon(v) &=\{1,\ldots,k-j+1\}\cup \{x+k-j+1\mid \varepsilon_x\in\Gamma_j\} \tag{*}
		    \intertext{Hence, from Equation~\eqref{eq def quot tree},}
		    U'_{j}&={t'}_{j-1}^{-1}U'_{j}\circ(t'_{j-1},\Delta_j)\\
		    &=U'_{j-1}\circ(t'_{j-1},\Delta_j)
		    \intertext{Since $U'_{j-1}={t'}_j^{-1}U'_j$, $U'_{j-1}$ is the only set satisfying $\forall v\in U'_{j-1}$}
		    u&=( v\circ(t'_{j-1},\Delta_j))\circ(t_j,\ldots,t_k,\Gamma_j) \tag{**}\\
		    \mathrm{Ind}_\varepsilon(v)&=\mathrm{Ind}_\varepsilon({t'}_{j-1}^{-1}U'_j)\\
		    &=\{1\}\cup\{2,\ldots,k-j+2\}\\
		    &\qquad\cup \{x+k-j+2\mid \varepsilon_x\in\Gamma_j\}\setminus\mathrm{Ind}_\varepsilon(t'_{j-1})\\
		    &=\{1\}\cup\{2,\ldots,k-j+2\}\\
		    &\qquad\cup \{x+k-j+2\mid \varepsilon_x\in\Gamma_j\}\setminus\mathrm{Ind}_\varepsilon(\mathrm{Inc}_\varepsilon(k-j+1,t_{j-1}))\\
		      &=\{1\}\cup\{2,\ldots,k-j+2\}\cup \{x+k-j+2\mid \varepsilon_x\in\Gamma_{j-1}\}\\
		      &=\{1\ldots,k-(j-1)+1\}\cup \{x+k-(j-1)+1\mid \varepsilon_x\in\Gamma_{j-1}\}
		      \intertext{From Equation~\eqref{eq calc ind compo} and (*), (**) becomes}
		      u&=
		        v\circ(t'_{j-1}\circ  (\varepsilon_{x_{z}})_{1\leq z\leq n\land x_z\in\mathrm{Ind}_\varepsilon(t_{j-1})},
		        \varepsilon_1\circ t_j,\ldots,\varepsilon_{k-j+1}\circ t_k,\\
		        &\qquad
		        (\varepsilon_{x_z+k-j+1}\circ \varepsilon_{x_z})_{x_z\in\Gamma_j \land x_z\notin\mathrm{Ind}_\varepsilon(t_{j-1})}   
		        )\\
		      &=v\circ(t_{j-1},
		      t_j,\ldots,t_k,  
		        \Gamma_{j-1} 
		      ) 
		  \end{align*}
		  \item 
		  Let $U'=f^{-1}(U'_1)$. Then, from Equation~\eqref{eq def quot tree}, $U'$ is the only set satisfying
		  \begin{align*}
		    U'_1&=U'\circ(f,\mathrm{Inc}_\varepsilon(k,\Gamma_1))
		    \intertext{Then, from previous point, $U'$ is the only set satisfying $\forall v\in U'$}
		    u&=(v\circ(f,\mathrm{Inc}_\varepsilon(k,\Gamma_1)))\circ (t_1,\ldots,t_k,\Gamma_1)\\
		    &=v\circ(f(t_1,\ldots,t_k),\Gamma_1)\\
		    \mathrm{Ind}_\varepsilon(v)&=\mathrm{Ind}_\varepsilon(f^{-1}U'_1)\\
		     &=\{1\}\cup \{x+k+1\mid \varepsilon_x\in\Gamma_1\}
		    \intertext{As a direct consequence, the set $U''$ defined by}
		    U''&=U'\circ (\varepsilon_1,(\varepsilon_{x_z+1})_{1\leq z\leq n\land x_z \in\Gamma_1})
		    \intertext{is the only set satisfying $\forall v\in U''$}
		    u&=v\circ(f(t_1,\ldots,t_k),\Gamma_1)\\
		    \mathrm{Ind}_\varepsilon(v)&=\{1\}\cup \{x+1\mid \varepsilon_x\in\Gamma_1\}
		    \intertext{that is by definition the set $t^{-1}(u)$.}
		  \end{align*}
		\end{enumerate}  
	  \cqfd
	\end{proof}	
	
	\noindent\textbf{Lemma~\ref{lem quot tree cdotb}.}
    \emph{Let $\Sigma$ be a graded alphabet.
    Let $t$ be a $k$-ary tree in $T_\Sigma$ and $L$ be a $0$-homogeneous language.
    Let $\alpha$ be a symbol in $\Sigma$ and $b$ be a symbol in $\Sigma_0$.
    Then:}
    \begin{align*}  
      \alpha^{-1}(t\cdot_b L)&= 
        \begin{cases}
          (b^{-1}(t)\cdot_b L) \circ_1 b^{-1}(L) & \text{ \emph{if} }\alpha=b,\\
          \alpha^{-1}(t)\cdot_b L \cup (b^{-1}(t)\cdot_b L) \circ_1 \alpha^{-1}(L) & \text{ \emph{if} }\alpha\in\Sigma_0\setminus\{b\},\\
          \alpha^{-1}(t)\cdot_b L & \text{\emph{otherwise.}}
        \end{cases}
    \end{align*}
  \begin{proof}
    If there is no occurrence of $b$ in $t$, then $t\cdot_b L=t$ and therefore
      \begin{align*}
        \alpha^{-1}(t\cdot_b L)&=
          \begin{cases}
            \emptyset= (b^{-1}(t)\cdot_b L) \circ_1 b^{-1}(L) & \text{ if }\alpha=b,\\
            \alpha^{-1}(t)\cdot_b L & \text{otherwise.}
          \end{cases}
      \end{align*}
    Otherwise let us proceed by induction over $t$.
    \begin{enumerate}
	    \item If  $t=b$, then $t\cdot_b L=L$ and 
	      \begin{align*}
	        \alpha^{-1}(t\cdot_b L)=\alpha^{-1}(L)=\varepsilon_1 \circ_1 \alpha^{-1}(L)=(b^{-1}(b)\cdot_b L) \circ_1 \alpha^{-1}(L)
	      \end{align*}
	    \item Consider that $t=f(t_1,\ldots,t_n)$ with $f\in\Sigma_n$, $n>0$. Then
	      \begin{align*}
	        \alpha^{-1}(t\cdot_b L)&=\alpha^{-1}(f(t_1\cdot_b L,\ldots,t_n\cdot_b L))\\
	        &=\bigcup_{1\leq l\leq n} f((t'_j)_{1\leq j\leq l-1}, \alpha^{-1}(t_l\cdot_b L),(t'_j)_{l+1\leq j\leq n}) 
	        \quad \text{(Proposition~\ref{prop calc quot symb})}
	        \intertext{where $t'_j=\mathrm{Inc}_\varepsilon(1,t_j\cdot_b L)$}
	        \intertext{By Induction hypothesis}
	        \alpha^{-1}(t\cdot_b L)&=\bigcup_{1\leq l\leq n} f((t'_j)_{1\leq j\leq l-1}, L'',(t'_j)_{l+1\leq j\leq n})
	        \intertext{where}
	        L''&=
	          \begin{cases}
	            (b^{-1}(t_l)\cdot_b L) \circ_1 b^{-1}(L) & \text{ if }\alpha=b,\\
	            \alpha^{-1}(t_l)\cdot_b L \cup (b^{-1}(t_l)\cdot_b L) \circ_1 \alpha^{-1}(L) & \text{ if }\alpha\in\Sigma_0\setminus\{b\},\\
	            \alpha^{-1}(t_l)\cdot_b L & \text{otherwise,}
	          \end{cases}
	      \end{align*}
	      \begin{enumerate}
	        \item Suppose that $L''=(b^{-1}(t_l)\cdot_b L) \circ_1 \alpha^{-1}(L)$. Then:
			      \begin{align*}
			        \alpha^{-1}(t\cdot_b L)&= \bigcup_{1\leq l\leq n} f((t'_j)_{1\leq j\leq l-1}, (b^{-1}(t_l)\cdot_b L) \circ_1 \alpha^{-1}(L),(t'_j)_{l+1\leq j\leq n})
			        \intertext{Since there is no occurrence of $\varepsilon_1$ in any $t'_j$:}
			         &= (\bigcup_{1\leq l\leq n} f((t'_j)_{1\leq j\leq l-1}, (b^{-1}(t_l)\cdot_b L) ,(t'_j)_{l+1\leq j\leq n}))\circ_1 \alpha^{-1}(L)\\
			         &= ((\bigcup_{1\leq l\leq n} f((\mathrm{Inc}_\varepsilon(1,t_j))_{1\leq j\leq l-1}, b^{-1}(t_l) ,(\mathrm{Inc}_\varepsilon(1,t_j))_{l+1\leq j\leq n}))\cdot_b L)\circ_1 \alpha^{-1}(L)\\
			         &= (b^{-1}(t)\cdot_b L)\circ_1 \alpha^{-1}(L) 
			         \quad \text{(Proposition~\ref{prop calc quot symb})}
			      \end{align*}
	        \item Suppose that $L''=\alpha^{-1}(t)\cdot_b L$. Then:
			      \begin{align*}
			        \alpha^{-1}(t\cdot_b L)&= \bigcup_{1\leq l\leq n} f((t'_j)_{1\leq j\leq l-1}, \alpha^{-1}(t)\cdot_b L,(t'_j)_{l+1\leq j\leq n})\\
			        &= (\bigcup_{1\leq l\leq n} f((\mathrm{Inc}_\varepsilon(1,t_j))_{1\leq j\leq l-1}, \alpha^{-1}(t),(\mathrm{Inc}_\varepsilon(1,t_j))_{l+1\leq j\leq n}))\cdot_b L\\
			        &= \alpha^{-1}(t)\cdot_b L 
			        \quad \text{(Proposition~\ref{prop calc quot symb})}
			      \end{align*}
			    \item If $L''=\alpha^{-1}(t_l)\cdot_b L \cup (b^{-1}(t_l)\cdot_b L) \circ_1 \alpha^{-1}(L)$, following the two previous items:
			      \begin{align*}
			        \alpha^{-1}(t\cdot_b L)&= \bigcup_{1\leq l\leq n} f((t'_j)_{1\leq j\leq l-1}, L'',(t'_j)_{l+1\leq j\leq n})\\
			         &=  \alpha^{-1}(t)\cdot_b L\cup(b^{-1}(t))\cdot_b L)\circ_1 \alpha^{-1}(L) 
			      \end{align*}
			  \end{enumerate}   
		\end{enumerate}
    \cqfd
  \end{proof} 
  
  \noindent\textbf{Lemma~\ref{lem quot tree compos}.}
    \emph{Let $\Sigma$ be a graded alphabet.
    Let $t$ be a $k$-ary tree with $\mathrm{Ind}_\varepsilon(t)=\{j_1,\ldots,j_k\}$  and $t_1,\ldots,t_k$ be $k$ trees.
    Let $\alpha$ be a symbol in $\Sigma_n$.
    Then:}
    \begin{align*} 
      \alpha^{-1}(t\circ(t_1,\ldots,t_k))&=
          \bigcup_{1\leq j\leq k} t\circ((\mathrm{Inc}_\varepsilon(1,t_l))_{1\leq l\leq j-1},\alpha^{-1}(t_j),(\mathrm{Inc}_\varepsilon(1,t_l))_{j+1\leq l\leq k}) \cup X
    \end{align*}
    where
    \begin{align*} 
          X &=
            \begin{cases}
               \alpha((\varepsilon_{j_{p_l}})_{1\leq l\leq n})^{-1}(t) \circ(\varepsilon_1,(\mathrm{Inc}_\varepsilon(1,t_l))_{1\leq l\leq k\mid\forall j, l\neq p_j}) & \text{ \emph{if} }\forall 1\leq l\leq n,  \exists 1\leq p_l\leq k, t_{p_l}= \varepsilon_l \\
               \emptyset & \text{ \emph{otherwise}.}
            \end{cases}  
    \end{align*}
  \begin{proof}
    By induction over the structure of $t$.
    Let us set $t'=t\circ(t_1,\ldots,t_k)$.
    \begin{enumerate}
      \item Consider that $t=f(\varepsilon_{j_{i_1}},\ldots,\varepsilon_{j_{i_k}})$ with $f\in\Sigma_k$.
      Then $t'=f(t_{i_1},\ldots,t_{i_k})$.
      \begin{enumerate}
        \item Consider that $\forall 1\leq l\leq n, \exists 1\leq p_l\leq n, t_{p_l}= \varepsilon_l$.
        If $t_{i_z}\neq \varepsilon_z$ for some $1 \leq z\leq n$ then according to Proposition~\ref{prop calc quot symb}, $\alpha^{-1}(t')=\emptyset$.
        Hence consider that $t_{i_z}=\varepsilon_z$ $\forall 1\leq z\leq n$. 
        Then $\forall 1\leq z\leq k$, $i_z=p_z$ and $\alpha^{-1}(t')=\{\varepsilon_1\}$.
        Furthermore $\alpha((\varepsilon_{j_{p_l}})_{1\leq l\leq k})^{-1}(t)=\alpha((\varepsilon_{j_{i_l}})_{1\leq l\leq k})^{-1}(t)=t^{-1}(t)=\{\varepsilon_1\}$.
        \item Otherwise, From Proposition~\ref{prop calc quot symb},
        \begin{align*}
          \alpha^{-1}(t')& =\bigcup_{1\leq l\leq k} f((\mathrm{Inc}_\varepsilon(1,t_{i_z})_{1\leq z \leq l-1}),\alpha^{-1}(t_{i_l}),(\mathrm{Inc}_\varepsilon(1,t_{i_z})_{l+1\leq z \leq k}))\\
          & =\bigcup_{1\leq l\leq k} f(\varepsilon_{j_{i_1}},\ldots,\varepsilon_{j_{i_k}})\circ ((\mathrm{Inc}_\varepsilon(1,t_{z})_{1\leq z \leq l-1}),\alpha^{-1}(t_{l}),(\mathrm{Inc}_\varepsilon(1,t_{z})_{l+1\leq z \leq k}))\\
          & =\bigcup_{1\leq l\leq k} t \circ ((\mathrm{Inc}_\varepsilon(1,t_{z})_{1\leq z \leq l-1}),\alpha^{-1}(t_{l}),(\mathrm{Inc}_\varepsilon(1,t_{z})_{l+1\leq z \leq k}))
        \end{align*}
      \end{enumerate} 
      \item Consider that $t=f(u_1,\ldots,u_k)$. Then from Equation~\eqref{eq calc ind compo}, $t'=f((u_l\circ T_l)_{1\leq l\leq k})$, with $T_l=(t_z)_{1\leq z\leq k\land j_z\in\mathrm{Ind}_\varepsilon(u_l)}$.
      From Proposition~\ref{prop calc quot symb},
      \begin{align*}
        \alpha^{-1}(t')&=\bigcup_{1\leq j\leq k}f((u'_z)_{1\leq z\leq j-1}, \alpha^{-1}(u_j\circ T_j),(u'_z)_{j+1\leq z\leq n})
        \intertext{ with $u'_l=\mathrm{Inc}_\varepsilon(1,u_l\circ T_l)$.}
        \intertext{Let us set $T_j=(t_{j,1},\ldots,t_{j,r_j})$ and  $w_j=u_j \circ T_j$}
        \intertext{By induction hypothesis,}
         \alpha^{-1}(w_j)& =
           \bigcup_{1\leq z\leq r_j} u_j\circ((\mathrm{Inc}_\varepsilon(1,t_{j,l}))_{1\leq l\leq z-1},\alpha^{-1}(t_{j,z}),(\mathrm{Inc}_\varepsilon(1,t_{j,l}))_{z+1\leq l\leq r_j}) \cup X_j\\
        \intertext{with}
           X_j&=
            \begin{cases}
               \alpha((\varepsilon_{j_{p_l}})_{1\leq l\leq n})^{-1}(u_j) \circ(\varepsilon_1,(\mathrm{Inc}_\varepsilon(1,t_{j,l}))_{1\leq l\leq r_j\mid\forall z, l\neq p_z})) & \text{if} \forall 1\leq l\leq n,\\
               & \exists 1\leq p_l\leq r_j, t_{j,p_l}=\varepsilon_l.\\
               \emptyset & \text{ otherwise.}
            \end{cases}
        \intertext{Hence}
        \alpha^{-1}(t')&=\bigcup_{1\leq j\leq k}f(
          Y_1 , 
          Y_2 ,
          Y_3) 
          \cup V\\
        \intertext{with}
          Y_1 &= (u'_z)_{1\leq z\leq j-1}\\
          Y_2 &= \bigcup_{1\leq z\leq p_j} u_j\circ((\mathrm{Inc}_\varepsilon(1,t_{j,l}))_{1\leq l\leq z-1},\alpha^{-1}(t_{j,z}),(\mathrm{Inc}_\varepsilon(1,t_{j,l}))_{z+1\leq l\leq k})\\
          Y_3 &= (u'_z)_{j+1\leq z\leq k})\\
          V &=
            \bigcup_{1\leq j\leq k}f((u'_z)_{1\leq z\leq j-1},X_j,(u'_z)_{j+1\leq z\leq k})\\     
        \intertext{Therefore, since $u'_l=\mathrm{Inc}_\varepsilon(1,u_l\circ T_l)=u_l\circ \mathrm{Inc}_\varepsilon(1, T_l)$}
        \alpha^{-1}(t')&=A \cup B
        \intertext{with}
        A&=\bigcup_{1\leq j\leq k}f(u_1,\ldots,u_k) \circ (\mathrm{Inc}_\varepsilon(1,t_l)_{1\leq l\leq j-1},\alpha^{-1}(t_j),\mathrm{Inc}_\varepsilon(1,t_l)_{j+1\leq l\leq k})\\
        &=\bigcup_{1\leq j\leq k} t \circ (\mathrm{Inc}_\varepsilon(1,t_l)_{1\leq l\leq j-1},\alpha^{-1}(t_j),\mathrm{Inc}_\varepsilon(1,t_l)_{j+1\leq l\leq k})
        \intertext{and if $\exists 1\leq j\leq k,\forall 1\leq l\leq n, \exists 1\leq p_l\leq r_j, t_{j,p_l}=\varepsilon_l$}
        B&=\bigcup_{1\leq j\leq k} f(\mathrm{Inc}_\varepsilon(1,u'_l)_{1\leq l\leq j-1},\alpha((\varepsilon_{j_{p_l}})_{1\leq l\leq n})^{-1}(u_j),\mathrm{Inc}_\varepsilon(1,u'_l)_{j+1\leq l\leq j-1})\circ\\
        & \quad (\varepsilon_1,(\mathrm{Inc}_\varepsilon(1,t_l))_{1\leq l\leq k\mid\forall z, l\neq p_z})) \\
        &=\alpha((\varepsilon_{j_{p_l}})_{1\leq l\leq n})^{-1}(f(u_1,\ldots,u_k))\circ(\varepsilon_1,(\mathrm{Inc}_\varepsilon(1,t_l))_{1\leq l\leq k\mid\forall z, l\neq p_z})) 
        \intertext{otherwise}
        B&=\emptyset
        \intertext{Since there is at most one $j$ such that $X_j\neq\emptyset$ and if there is none, it implies that}
        B&=\emptyset
        \intertext{Then the formula holds.}  
      \end{align*}
    \end{enumerate}
    \cqfd
  \end{proof}  
  
  \noindent\textbf{Proposition~\ref{prop bot up quot star rond}.}
    \emph{Let $\Sigma$ be a graded alphabet.
    Let $L$ be a $1$-homogeneous language.
    Let $\alpha$ be a symbol in $\Sigma_0$.
    Then:}
    \begin{align*} 
      \alpha^{-1}(L^{\circledast})&=
        \begin{cases}
          (L^\circledast\circ (\alpha^{-1}(L)))\circ(\varepsilon_1, \mathrm{Inc}_{\varepsilon}(1,L^\circledast)) & \text{ \emph{if} }\alpha\in\Sigma_0,\\
          (L^\circledast\circ (\alpha^{-1}(L))) & \text{ \emph{otherwise}.}
        \end{cases}
    \end{align*}
  \begin{proof}
	  By definition, for any integer $n$,
	    \begin{align*} 
	      \alpha^{-1}(L^{n+1_\circ})&=L\circ L^{n_\circ} \cup L^{n_\circ}
        \intertext{Following Proposition~\ref{prop quot lang circ}:}
	      \alpha^{-1}(L\circ L^{n})&= 
	        \begin{cases}
	          L\circ \alpha^{-1}(L^{n_\circ}) \cup \alpha^{-1}(L)\circ(\varepsilon_1,\mathrm{Inc}_\varepsilon(1,L^{n_\circ})) & \text{ if } \alpha\in\Sigma_0,\\
	          L\circ \alpha^{-1}(L^{n_\circ}) \cup \alpha^{-1}(L) & \text{ otherwise.}
	        \end{cases}
	      \intertext{Hence,}
	      \alpha^{-1}(L^{n+1_\circ})&= 
	        \begin{cases}
	          \alpha^{-1}(L^{n_\circ})\cup L\circ \alpha^{-1}(L^{n_\circ}) \cup \alpha^{-1}(L)\circ(\varepsilon_1,\mathrm{Inc}_\varepsilon(1,L^{n_\circ})) & \text{ if } \alpha\in\Sigma_0,\\
	          \alpha^{-1}(L^{n_\circ})\cup L\circ \alpha^{-1}(L^{n_\circ}) \cup \alpha^{-1}(L) & \text{ otherwise.}
	        \end{cases}
	      \intertext{Therefore,}
	      \alpha^{-1}(L^{\circledast})&= \bigcup_{j\geq 0} \alpha^{-1}(L^{j_\circ})\\
	      &=\bigcup_{j,k\geq 0}
	        \begin{cases}
	        L^{j_\circ} \circ \alpha^{-1}(L)\circ(\varepsilon_1,\mathrm{Inc}_\varepsilon(1,L^{k_\circ})) & \text{ if } \alpha\in\Sigma_0,\\
	        L^{j_\circ} \circ \alpha^{-1}(L) & \text{ otherwise.}
	        \end{cases}\\
	      &=
	        \begin{cases}
	        L^{\circledast} \circ \alpha^{-1}(L)\circ(\varepsilon_1,\mathrm{Inc}_\varepsilon(1,L^{\circledast})) & \text{ if } \alpha\in\Sigma_0,\\
	        L^{\circledast} \circ \alpha^{-1}(L) & \text{ otherwise.}
	        \end{cases}
		  \end{align*}
    \cqfd
  \end{proof}

  \noindent\textbf{Proposition~\ref{prop bot up quot star symb}.}
    \emph{Let $\Sigma$ be a graded alphabet.
    Let $L$ be a $0$-homogeneous language.
    Let $\alpha$ be a symbol in $\Sigma$ and $b$ be a symbol in $\Sigma_0$.
    Then:}
    \begin{align*} 
      \alpha^{-1}(L^{*_b})&=
        \begin{cases}
          (b^{-1}(L))^\circledast\cdot_b L^{*_b} & \text{ \emph{if} }\alpha=b,\\
          ((b^{-1}(L))^\circledast \circ (\alpha^{-1}(L))) \cdot_b L^{*_b} & \text{\emph{otherwise}.}
        \end{cases}
    \end{align*}
  \begin{proof}
    By definition, for any integer $n$,
      \begin{align*}
	      \alpha^{-1}(L^{n+1_b})&=\alpha^{-1}(L^{n_b}\cup L\cdot_b L^{n_b} )\\
	      &=\alpha^{-1}(L^{n_b})\cup \alpha^{-1}(L\cdot_b L^{n_b} )
	      \intertext{Moreover, according to Proposition~\ref{prop quot cdotb lang},}	
	      \alpha^{-1}(L\cdot_b L^{n_b} )&=
        \begin{cases}
          (b^{-1}(L)\cdot_b L^{n_b}) \circ b^{-1}(L^{n_b}) & \text{ if }\alpha=b,\\
          \alpha^{-1}(L)\cdot_b L^{n_b} \cup (b^{-1}(L)\cdot_b L^{n_b}) \circ \alpha^{-1}(L^{n_b}) & \text{ if }\alpha\in\Sigma_0\setminus\{b\},\\
        \end{cases}
        \intertext{Hence, since by definition, $\varepsilon_1$ is in $L^{0_b}\subset L^{n_b}$,}
	       \alpha^{-1}(L^{n+1_b})
	      &=
	        \begin{cases}
	         \{\varepsilon_1\} \cup b^{-1}L^{n_b} \cup (b^{-1}(L)\cdot_b L^{n_b})\circ(b^{-1}(L^{n_b}))& \text{ if }\alpha=b,\\
	           \alpha^{-1}(L)\cdot_b L^{n_b} \cup \alpha^{-1}(L^{n_b}) \cup (b^{-1}(L)\cdot_b L^{n_b})\circ(\alpha^{-1}(L^{n_b})) & \text{ otherwise.}
	        \end{cases}
	        \intertext{As a direct consequence,}
	      \alpha^{-1}(L^{*_b})&=\bigcup_{j\geq 0} \alpha^{-1}(L^{j_b})\\
	        &=\bigcup_{j\geq 1, p_j\geq\cdots\geq 1 0}
	        \begin{cases}
	         \{\varepsilon_1\} \cup  (b^{-1}(L)\cdot_b L^{{p_j}_b})\circ \cdots \circ (b^{-1}(L)\cdot_b L^{{p_1}_b})& \text{ if }\alpha=b,\\
	          (b^{-1}(L)\cdot_b L^{{p_j}_b})\circ \cdots \circ (b^{-1}(L)\cdot_b L^{{p_2}_b})\circ (\alpha^{-1}(L)\cdot_b L^{{p_1}_b}) & \text{ otherwise,}
	        \end{cases}\\
	        &=\bigcup_{j\geq 0, k\geq 1}
	        \begin{cases}
	         \{\varepsilon_1\} \cup  \overbrace{(b^{-1}(L)\cdot_b L^{{j}_b})\circ \cdots \circ (b^{-1}(L)\cdot_b L^{{j}_b})}^{\text{$k$ times}}& \text{ if }\alpha=b,\\
	         \underbrace{(b^{-1}(L)\cdot_b L^{{j}_b})\circ \cdots \circ (b^{-1}(L)\cdot_b L^{{j}_b})}_{\text{$k-1$ times}} \circ (\alpha^{-1}(L)\cdot_b L^{{j}_b}) & \text{ otherwise,}
	        \end{cases}\\
	        &=\bigcup_{j\geq 0, k\geq 1}
	        \begin{cases}
	         \{\varepsilon_1\} \cup  (b^{-1}L)^{k_\circ}\cdot_b L^{j_b}& \text{ if }\alpha=b,\\
	         ((b^{-1}L)^{k-1_\circ}\circ (\alpha^{-1}(L))\cdot_b L^{j_b}& \text{ otherwise,}
	        \end{cases}\\
	        &=
	        \begin{cases}
	          (b^{-1}(L))^\circledast\cdot_b L^{*_b} & \text{ if }\alpha=b,\\
	          ((b^{-1}(L))^\circledast \circ (\alpha^{-1}(L))) \cdot_b L^{*_b} & \text{otherwise.}\\
	        \end{cases}
	    \end{align*}
    \cqfd
  \end{proof}
  
  \noindent\textbf{Proposition~\ref{prop lien quot lang et lang haut}.}
    \emph{Let $A=(\Sigma,Q,F,\delta)$ be a automaton.
    Then, for any tree $t$ in $T_{\Sigma}$, it holds:}
    \begin{align*}
      t^{-1}(L(A))&=
        \bigcup_{q\in \Delta(t)} L^{q}(A)
    \end{align*}
  \begin{proof}
    By definition,
    \begin{align*}
      t^{-1}(L(A))&=\{t'\in T_{\Sigma,1}\mid t'\circ t\in L(A)\}\\   
      &=\{t'\in T_{\Sigma,1}\mid \Delta(t',\Delta(t))\cap F\neq\emptyset\}\\
      &= \bigcup_{q\in \Delta(t)} L^{q}(A) 
    \end{align*}
    \cqfd
  \end{proof} 
  
  \noindent\textbf{Theorem~\ref{thm lien card aut card quot}.}
    \emph{Let $A=(\Sigma,Q,F,\delta)$ be a deterministic tree automaton.
    Then:}
    \begin{align*}
      \mathrm{Card}(\{t^{-1}(L(A))\mid t\in T_{\Sigma}\}) \leq \mathrm{Card}(Q)
    \end{align*}
  \begin{proof}
    According to Proposition~\ref{prop lien quot lang et lang haut}, for any tree $t$ in $T_\Sigma$,
    \begin{align*}
      t^{-1}(L(A))&= \bigcup_{q\in \Delta(t)} L^{q}(A)
      \intertext{Hence, since $A$ is deterministic, for any tree $t$ in $T_\Sigma$,}
      \mathrm{Card}(\Delta(t))&\leq 1
      \intertext{And}
      t^{-1}(L(A))&\in \bigcup_{q\in Q} \{L^{q}(A)\}
      \intertext{Since}
      \mathrm{Card}(\bigcup_{q\in Q} \{L^{q}(A)\})\leq \mathrm{Card}(Q)
      \intertext{Theorem holds.}
    \end{align*}
    \cqfd
  \end{proof}
  
  \noindent\textbf{Proposition~\ref{prop aut min reco l}.} 
    \emph{Let $L$ be a tree language in $\mathcal{L}(\Sigma)_0$.
    Then $L(A_L)=L$.}
  \begin{proof}
    Let us show by induction over a tree $t$ in $T_\Sigma$ that $\Delta(t)=\{t^{-1}(L)\}$.
    \begin{align*}
      \Delta(f(t_1,\ldots,t_n))&=\delta(f,\Delta(t_1),\ldots,\Delta(t_n))\\
      &=\delta(f,\{t_1^{-1}(L)\},\ldots,t_n^{-1}(L))\\
      &=\{(f(t_1,\ldots,t_n))^{-1}(L)\}\\
    \end{align*}
    Consequently, 
    \begin{align*}
      t\in L(A_L) &\Leftrightarrow \Delta(t)\cap F\neq\emptyset\\
      &\Leftrightarrow (t^{-1}(L))\in F\\
      &\Leftrightarrow \varepsilon_1\in (t^{-1}(L))\\
      &\Leftrightarrow t\in L
    \end{align*}
    \cqfd
  \end{proof}
  
  \noindent\textbf{Proposition~\ref{prop morphism from dfa to min}.}
    \emph{Let $A$ be an accessible deterministic tree automaton.
    Let $\phi$ be the function that associates to any state $q$ in $Q$ the language $L^q(A)$.
    Then $\phi$ a morphism from $A$ to $A_{L(A)}$.}
  \begin{proof}
    Let $A=(\Sigma,Q,F,\delta)$.
    From Proposition~\ref{prop lien quot lang et lang haut}, for any tree $t$ in $L_q(A)$, $L^q(A)=t^{-1}(L(A))$. Hence, for any $q$ a tree $t$ in $L_q(A)$, $\phi(q)=t^{-1}(L(A))$. 
    
    Suppose that $q$ is final. Then for any tree $t$ in $L_q(A)$, $t\in L(A)$. Hence $\varepsilon_1\in t^{-1}(L(A))$ and therefore $ t^{-1}(L(A))$ is final in $A_L$.
    
    Consider a transition $(q,f,q_1,\ldots,q_n)$ in $\delta$. By definition, $f(L_{q_1}(A),\ldots,L_{q_n}(A))\subset L_q(A)$. Consequently, for any $n$ trees $(t_1,\ldots,t_n)$ in $(L_{q_1}(A),\ldots,L_{q_n}(A))$, $f(t_1^{-1}(L(A)),\ldots,$ $t_n^{-1}(L(A)))\subset L_q(A)$ and $f(t_1,\ldots,t_n)$ is in $ L_q(A)$. Therefore $\phi(q)=f(t_1,\ldots,t_n)^{-1}(L(A))$, and since by construction $(f(t_1,\ldots,t_n)^{-1}(L(A)),f,t_1^{-1}(L(A)),\ldots,t_n^{-1}(L(A)))$ is a transition in $A_L$, for any transition $(q,f,q_1,\ldots,q_k)$ in $\delta$, $(\phi(q),f,\phi(q_1),\ldots,\phi(q_k))$ is a transition in $A_L$.
    \cqfd
  \end{proof}
  
\end{document}